\documentclass[sigconf]{acmart}

\AtBeginDocument{%
  \providecommand\BibTeX{{%
    \normalfont B\kern-0.5em{\scshape i\kern-0.25em b}\kern-0.8em\TeX}}}

\copyrightyear{2020}
\acmYear{2020}
\setcopyright{iw3c2w3}
\acmConference[WWW '20]{Proceedings of The Web Conference 2020}{April 20--24, 2020}{Taipei, Taiwan}
\acmBooktitle{Proceedings of The Web Conference 2020 (WWW '20), April 20--24, 2020, Taipei, Taiwan}
\acmPrice{}
\acmDOI{10.1145/3366423.3380180}
\acmISBN{978-1-4503-7023-3/20/04}

\usepackage{amssymb,amsmath,amsthm,dsfont,bm,mathtools}
\usepackage{color,graphicx}
\graphicspath{{figures/}}
\usepackage{tabularx}

\newcommand{\loss}{{L}}
\newcommand{\yhat}{{\hat{y}}}

\newcommand{\tg}{{\tau}}
\newcommand{\lblfunc}{{\ell}}
\newcommand{\iter}{{h}}

\newcommand{\beq}{\begin{equation}}
\newcommand{\eeq}{\end{equation}}
\newcommand{\bal}{\begin{align}}
\newcommand{\eal}{\end{align}}
\newcommand\expect[2]{\mathbb{E}_{#1}{[ {#2} ]}}

\newcommand{\1}[1]{\mathds{1}{\{{#1}\}}}

\newcommand{\naive}{{na\"{\i}ve}}

\begin{document}

\title[A Kernel of Truth]{A Kernel of Truth: Determining Rumor Veracity \\ on Twitter by Diffusion Pattern Alone}

\author{Nir Rosenfeld}
\affiliation{%
	\institution{Harvard University}
	\department{School of Engineering \\ and Applied Sciences}
}%
\email{nirr@seas.harvard.edu}
\authornote{These authors contributed equally to the paper and are ordered alphabetically}

\author{Aron Szanto}
\affiliation{%
	\institution{Harvard University}
	\department{School of Engineering \\ and Applied Sciences}
}%
\email{aron@seas.harvard.edu}
\authornotemark[1]
\authornote{This author is also a  Machine Learning Engineer at Kensho Technologies}
\authornote{Corresponding author}

\author{David C. Parkes}
\affiliation{%
	\institution{Harvard University}
	\department{School of Engineering \\ and Applied Sciences}
}%
\email{parkes@eecs.harvard.edu}
\authornote{This author is also an Affiliate of the Harvard Data Science Initiative}

\renewcommand{\shortauthors}{Szanto et al.}

\begin{abstract}

Recent work in the domain of misinformation detection has leveraged rich signals in the text and user identities associated with content on social media.
But text can be strategically manipulated
and accounts reopened under different aliases,
suggesting that these approaches are inherently brittle.
In this work, we investigate an alternative modality that is naturally robust: the pattern in which information propagates. 
Can the veracity of an unverified rumor spreading online be discerned solely on the basis of its pattern of diffusion through the social network?

Using graph kernels to extract complex topological information from Twitter cascade structures, we train accurate predictive models that are blind to language, user identities, and time, demonstrating for the first time that such ``sanitized'' diffusion patterns are highly informative of veracity. 
Our results indicate that, with proper aggregation,
the collective sharing pattern of the crowd may reveal powerful signals of rumor truth or falsehood,
even in the early stages of propagation.

\end{abstract}




\maketitle


\section{Introduction}

Social media has become the world's dominant vector for information propagation. Over three billion people use social media platforms such as Twitter and Facebook, and more than two thirds of Americans consume their news primarily through social media \cite{numsocialmedia_kemp2017, socialmedianews_shearer_gottfried_2017}. As social discourse, business, and even government are increasingly conducted on social media platforms, understanding how information travels through online networks is key to the study of the future of the Web.

When considering the dynamics of information propagation on social media, two factors are most salient: what is being shared and how. The former relates to content -- \emph{What is being discussed? By whom, and in association with which emotions? Is the information reliable?} -- while the latter refers to the spatiotemporal patterns that the information being shared induces on the online social network. 

The dynamics of these patterns are complex: the path that information takes on social media depends on the aggregation of many individual decisions to share a piece of content. But each decision distills the user's personality, behavior, and  emotional response as they relate to a simple binary choice---whether to share the information or not.

In this work, we ask whether the pattern in which information spreads is indicative of its content. In particular, we consider the following question: if we observe only \emph{how} information propagates, what can we infer about \emph{what} is being propagated?
We take a machine learning approach to this problem,
and in order to establish that information dispersion is indicative of information content,
build predictive models of content from patterns of propagation. 

\begin{figure}[t!]
	\centering
	\includegraphics[width=1.0\linewidth]{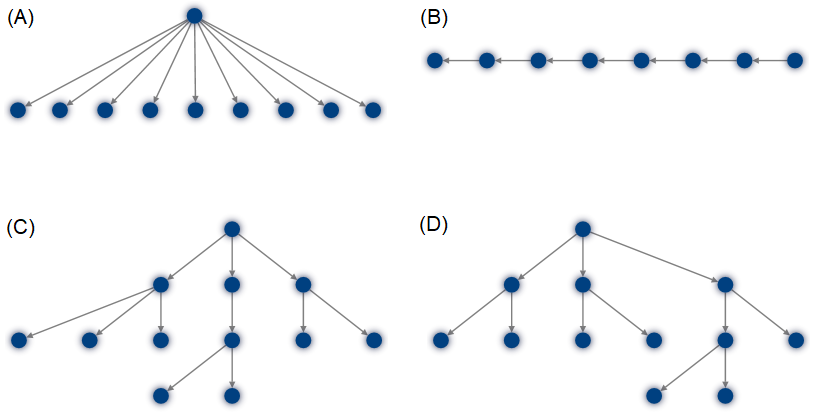}
	\caption{
		One of networks A and B describes how a wedding invitation propagates, while the other depicts the spread of a secret. Determining which is which is an easy task. In contrast, networks C and D both describe the spread of a rumor, one true and the other false. Can you guess which is which? In this paper we show that the truth or falsehood of rumors can be predicted just by observing their patterns of diffusion,
		without access to textual content or user identities,
		and before their veracity is public knowledge.}
	\label{fig:motivating_illustration}
\end{figure}

Of course, the relationship between content and propagation pattern can be quite simple. Individuals receiving a wedding invitation from a couple are unlikely to forward it further, resulting in a star-shaped path of information with the betrothed at the center. In contrast, consider the spread of a secret. If each party to the secret shares it with a single confidant, the pattern of propagation will resemble a chain. Accordingly, training a classifier to determine whether an information propagation pattern is associated with an invitation or a secret would be easy. The associated cascades would differ considerably under simple graph measures, such as size, breadth, depth, and branching factor. Other content types, however, may not be so distinct in their  patterns of diffusion (see Figure \ref{fig:motivating_illustration}).

We hypothesize that more opaque aspects of content can be accurately predicted from diffusion patterns alone. In this paper, we test this theory on the task of predicting rumor veracity from Twitter propagation patterns, also known as \emph{cascades}. Crucially, we restrict the information available to completely exclude any textual, temporal, or user-identifying data, leaving behind only the anonymous topology of the propagation patterns of rumors through Twitter. These patterns, modeled as simple directed graphs, reflect the collective behavior of users engaging with and differentially sharing rumors, and quite likely acting without verified knowledge of their truthfulness. We refer to these carefully curated cascades as \emph{sanitized cascades} and discuss their preparation in detail.

Our hypothesis that rumor veracity can be inferred from cascade topology arises from three observations from the literature. First, individual users react differently to true and false content \citep{pennycook2019fighting}.
Second, the collective opinions and actions of users can be aggregated to produce accurate forecasts, for example in prediction markets \citep{cowgill2015corporate}.
Third, in expectation, cascades surrounding true and false content differ in their patterns of propagation \citep{VosoughiRoyRumorsScience}.

Our results show that by combining appropriate representations of cascades and their substructures with careful aggregation of the collective behavior of users,
it is possible to discern rumor veracity, whereas simple metrics that capture global properties of cascades are too coarse to be useful.

We choose the domain of sanitized rumor cascades on Twitter because success therein evinces two particularly interesting conclusions.

First, evidence that rumor veracity can be predicted from diffusion patterns on Twitter suggests a surprising social interpretation:
although users spreading a rumor are likely unaware of its veracity at the time of sharing, the emergent, collective signal arising from the diffusion structure of the rumor through a crowd can, with proper aggregation, be informative of veracity. 

What makes the example of wedding invitation and secrets propagation uninteresting to study is that participating users are aware of the operative characteristic of the content and can adjust their behavior accordingly. In establishing that diffusion patterns on Twitter are informative of veracity even when users are locally uncertain,
we demonstrate how, with proper aggregation, weak individual signals can be combined
to produce useful predictions.
One of the unique contributions of our work is that we learn this aggregation from data.

Second, mitigating the spread of misinformation on social media is key to building a safer Web. To this end, validating our hypothesis on this task involves demonstrating a method for predicting a rumor's veracity from its diffusion pattern. Such a method seems likely to be more robust to interference by malicious producers of false content than does a model that relies on access to the content itself or to user identities associated with the rumor. While a single individual can fool content-based models by perturbing some text or whitewashing their identity, attacking a model that takes sanitized cascades as input requires changing the network topology, an undertaking that inherently demands concerted malicious action that spans many target users.

Our work draws on the same carefully curated data set as that used in \citet{VosoughiRoyRumorsScience}.
One of the key findings of that study is that false rumors, on average, spread deeper and faster through a social network. While their work provides valuable descriptive information, it reveals little about the predictive power of these statistical attributes with respect to rumor veracity.
Moreover, in the present paper we show that baseline models constructed to consider the statistics they extract---and show to be statistically significantly different, conditioned on rumor veracity---are not successful in predicting veracity.

While true and false rumors induce measurably different cascades, these disparities are not useful when trying to discriminate between information and misinformation. In light of this, we develop a model that extracts complex topological features from anonymous rumor cascades to make accurate predictions about veracity. In particular, we make use of \emph{graph kernels}, wherein graphs are embedded in a vector space, with dimensions corresponding to attributes of motif-like substructures \citep{Milo824Motif}. Here, a single Twitter rumor cascade is represented as a graph that the kernel maps to a point in the embedding space.

We introduce an efficient implementation of the Weisfeiler-Lehman graph kernel \citep{shervashidze2011weisfeiler} and use it to derive a topologically rich representation of rumor cascades on Twitter. Embedded cascades are then forwarded to a classifier that learns a relation between the embedding space and a binary label, namely rumor veracity. Ultimately, this technique embodies a model that can make accurate predictions about the veracity of a rumor solely on the basis of its sanitized cascade.

Before proceeding, it is important to stress what this work does \emph{not} intend to contribute. The model that we propose does not intend to monitor or detect fake news or to outperform benchmarks on misinformation classification. 
Methods for those tasks should be designed to utilize every data source available, including the rich content-based features that others have shown convincingly to be predictive of veracity--- the very same information that we remove via sanitization. Indeed, this work does not even necessarily concern fake news: we study rumors, and more specifically, rumors that have been determined to be worthy of investigation and publication by fact-checking websites.

Fundamentally, this paper explores an unstudied modality of prediction in which only the broad pattern of information diffusion is available to a model.


\subsection{Related Work}

Despite the relatively recent rise of fake news as a cultural fixture, there has long been academic interest in the propagation of true and false information through social media networks. \citet{friggeri2014rumor} describe the ways in which rumors spread on Facebook, finding that trust relationships play an important role. \citet{extra1zhou2007influence} describe theoretical properties of rumor propagation, while \citet{bakshy_role_2012} model information propagation and virality using field experiments to determine whether strong or weak ties between users are more responsible for the diffusion of novel information. \citet{gabielkov_social_2016} report results on Twitter in particular, studying different types of retweet cascade patterns. And \citet{extra7coletto2017motif} use graph motifs to model rumor controversy but stop short of predicting veracity.

There have also been several methods proposed to predict information veracity on social media platforms, but these have a different goal in mind,
and are intentionally designed to make use of content---linguistic, user-based, or image data ---in their input.
\citet{castillo_information_2011}, \citet{liu2015real}, \citet{ma2016detecting}, \citet{extra4Sampson}, and \citet{tacchini_like_2017} each use statistical models of article and tweet text, sometimes enriched with detailed user histories, to classify rumors after the fact as true or false; see also \citet{zhou2018fakesurvey} for a literature review of linguistic features studied for their predictiveness of disinformative content. 

Deep learning approaches include \citet{yu2017convolutional}, \citet{csiruchansky2017}, \citet{liu2018early}, and \citet{maetal2018rumor}, who combine linguistic features with news website data, user profile data, or the conversational stance of the content, using convolutional and recurrent neural networks to achieve good predictive performance on datasets from Twitter and its Chinese analog Weibo. \citet{Khattar2019mvaeWWW} integrate methods from computer vision to predict veracity on the basis of images and language. \citet{yang2019xfakeWWW} develop the XFake system for making and explaining predictions about news veracity, and \citet{MaGANWWW2019} propose an adversarial system based on textual content to disrupt the propagation of rumors, especially as enhanced by social network bots.

Some approaches combine content-based data with structural information about cascades. \citet{extra6wu2018tracing} propose a neural model that predicts rumor veracity based on cascade structure, but demands both user-identifying and temporal information as input.
\citet{extra2kwon2017rumor} develop a model that considers complex descriptive statistics about rumor cascades, but at the same time also requires knowledge of user-identifying follower relationships and performs roughly as good as chance even a week after a rumor is first published.

The above research is only tangential to the present paper,
as we purposefully focus only on topological information, and look to explore whether this in and of itself can be  predictive of veracity. Moreover, it is expressly not our intention to develop a fake news detector. Instead, we are motivated to explore a novel hypothesis as to whether rumor veracity can be inferred from sanitized cascade data alone, these cascades evolving before a fact checking event.
 
The study that is most related to our research is that of \citet{VosoughiRoyRumorsScience}, which characterizes the differential diffusion of true and false rumors through Twitter. Indeed, in this work we perform analysis on the same dataset. However, while they provide a descriptive study of rumor cascades and study metrics that differ significantly between true and false cascades, we show that these same metrics cannot be used to predict the veracity of a rumor. As far as we know, ours is the first work to perform predictive analytics on this important dataset, demonstrating that true and false rumors can be successfully differentiated by the shapes of their sanitized cascades. 

Graph kernels have been used for classification of networked objects in fields such as
computational biology \citep{ramon2003expressivity} and chemistry \citep{shervashidze2011weisfeiler},
but the literature on the use of graph kernels in social network classification is relatively small. One study develops {\em deep graph kernels}, which are graph kernel representations learned by neural networks, and applies them to predict Reddit sub-community interactions \citep{yanardag2015deep}. \citet{nikolentzos2017kernel} use a graph kernel based on a convolutional neural network that performs well on a synthetic dataset but has limited success on real-world social network datasets. Finally, \citet{extra3wu2015false} predict rumor veracity on Weibo using a graph kernel-based model, but their approach requires both rich textual data and user profile information. Moreover, though these studies involving graph kernels are methodologically relevant, none of them makes a comparison, as we do, to a baseline classifier that uses standard features extracted from the cascades. 

\section{Method}

\label{problem}
The task of interest in this work is to predict an attribute associated with a rumor, namely its veracity, having access only to the cascade that characterizes the rumor's propagation. We begin with a formal model of cascades and discuss
how they can be curated from data. We then describe a family of predictive models
suited for our task.

\subsection{Formal Setup}
\label{definition}
Let $G=(V,E)$ be a directed graph representing a social network (e.g., the Twitter follower graph), where $V$ is the set of nodes and $E$ is the set of edges.

Our input data comprise \emph{cascades}, objects describing the set of nodes involved in a single process of information sharing.
A \emph{cascade} $c = (V_c,E_c)$ is a connected, rooted digraph whose edges emanate outward from its root (otherwise known as an \emph{arborescence}).
Nodes $V_c \subseteq V$ indicate involved users
(with the root of $c$ corresponding to the individual initiating the process),
and $E_c \subseteq E$ are edges through which information flows.

Our data comprise cascades relating to rumors as they propagate through the Twitter social graph.
We denote by $r_c$ the rumor associated with $c$. Hence, an edge $(u,v) \in E_c$ indicates that $v \in V_c$ retweeted a (re)tweet from $u \in V_c$ that is discussing $r_c$.\footnote{Note that because retweets are broadcast, and because the data include only retweets (and not comments or likes), an edge $(u, v) \in E$ appears in $E_c$ only if both $u$ \emph{and} $v$ actively share the propagating content item.}
Each rumor may have multiple cascades related to it, but  each cascade relates only to one rumor. This will be important in designing our experimental evaluation scheme.

Each cascade $c$ is associated with a \emph{ground-truth label} $y \in \{0,1\}$, describing the truthfulness--- or veracity--- of $r_c$. Crucial to our study is that cascade $c$ does not directly encode any information related to rumor $r_c$ or ground truth $y$; $c$ has no access to the language associated with $r_c$, the user identities associated with $V_c$, or the temporal data associated with $E_c$. Any information about the content of rumor $r_c$ or its ground truth label $y$ can only be present, implicitly, in the topology of $c$.
We refer to these cascades as \emph{sanitized}. Details on the construction of such cascades are given in Section \ref{data_collection}.

To establish that cascades are \emph{informative} of rumorous content,
we seek to demonstrate that cascades are \emph{predictive} of veracity.
Our task, then, is to learn a function $f(c)$ mapping cascades $c$ to veracities $y \in \{0,1\}$.
We adopt the standard supervised learning paradigm and assume there is an unknown joint distribution $D$ over cascades and labels,
from which we observe a sample set $S=\{(c_j,y_j)\}_{j=1}^N$
of $N$ cascade-label pairs drawn i.i.d. from $D$.
Our learning objective is to train a classifier $f \in F$,
from some function class $F$, achieving high expected accuracy:
\begin{equation}
\expect{(c,y) \sim D}{\1{f(c)=y}}
= P_{(c,y) \sim D}(f(c)=y)
\end{equation}

which we achieve by minimizing an appropriate empirical loss:
\beq
\min_{f \in F} \sum_{j=1}^N \loss(y_j,f(c_j))
\eeq
where $\loss(y,\yhat)$ is a loss function
(e.g., the 0/1 loss, or some continuous proxy thereof)
and possibly using some form of regularization.

In the experiments we consider and evaluate several methods for classification,
each inducing a different function class $F$.

\subsection{Graph Kernels}
\label{modeling_approach}

Our main approach to predicting veracity from cascades uses graph kernels,
specifically an embodiment of the {\em Weisfeiler-Lehman} (WL) graph kernel \citep{shervashidze2011weisfeiler}.

Graph kernels are functions that compute inner products between vector representations of graphs.
One way to define a graph kernel is through its induced vector representation.\footnote{Graph kernels apply to graphs in general, but for consistency, we describe them in the context of cascades.}
Letting $\Omega$ be the set of cascades and $\mathcal{H}$ be a reproducing Hilbert space, consider a mapping $\phi: \Omega \rightarrow \mathcal{H}$.
A graph kernel $k: \Omega \times \Omega \rightarrow \mathbb{R}$ is a function
that first maps each graph into an embedded vector space via the mapping $\phi$,
and then computes an inner product within the embedded space:

\begin{equation}
k(c, c') = \langle\phi(c), \phi(c')\rangle_{\mathcal{H}}
\end{equation}

That is, for a cascade pair $(c, c')$, the kernel value is equal to their Hilbert-space inner product.

Hence, a graph kernel defines a similarity metric between two graphs, where similarity is measured in the embedded space; 
the larger the value of $k(c, c')$, the more similar $c$ and $c'$.
Different kernels induce different similarity metrics depending on the graph representations they consider.
Most graph kernels are based on an enumeration of certain graph substructures (e.g., paths, subgraphs, and random walks),
giving topologically rich (albeit combinatorially large) representations 
that capture intricate structural properties of graphs.

Graph kernels lie at the heart of many machine learning methods, as they offer
an expressive alternative to standard feature-engineered approaches;
rather than relying on handcrafted features relating to specific narrow properties,
graph kernels take the full shape of a graph into account,
incorporating both local and global information.
Due to the prohibitive size of representations, many methods handle features implicitly by working directly with kernel values $k(c,c')$.
Fortunately, due to our setting and the graph kernels we select, the
graph representations induced by the kernel are sparse (i.e., each $\phi(c)$ has relatively few non-zero entries). With proper consideration in implementation, then, these representations can be computed explicitly. This allows us to work directly with the graph embedding $\phi$ and consider any machine learning model that admits a sparse vector input (e.g., linear and tree-based models).

The literature on learning with graph inputs describes three overarching categories of graph kernels: random-walk kernels, shortest-paths kernels, and subtree kernels.
Since our data consists of cascades (in the form of arborescences),
and since cascades of real-world sharing patterns tend to be shallow (the median longest path across our cascades is four), kernels based on random walks and shortest paths are unlikely to offer sufficiently expressive representations of our inputs.
Thus, we choose to focus on subtree methods,
hypothesizing that these are particularly well-suited for capturing useful substructures of rumor cascades on social networks.
In particular,
our approach leverages the Weisfeiler-Lehman (WL) kernel \citep{shervashidze2011weisfeiler},
which we describe in detail next.

\subsection{The Weisfeiler-Lehman Kernel}
\label{model_wl}

The WL kernel is a subtree-based approach that measures the similarity of labeled graphs by iteratively comparing common node substructures, merging nodes by edge,
and then comparing again. It derives its name and underlying technique from the Weisfeiler-Lehman test of isomorphism between graphs \citep{wlOriginal}, which it applies sequentially to compute a metric for how close to (or far from) isomorphism two graphs are.

While the full WL procedure requires iterating until convergence,
in practice it is often useful to consider only a finite and predetermined number of iterations $\iter$, since this provides a means for controlling the size of the feature representation $\phi$.
One interpretation of the graph embedding $\phi$ induced by WL is that it represents a graph as a bag of motifs, similarly to how text can be represented as a bag of words
(indeed, our experiments show that methods known to work well for sparse text  also work well for WL graph representations).
Thus, each dimension of $\phi(c)$ corresponds to a motif,
and the entire vector $\phi(c)$ describes the collection (i.e., counts) of motifs in the input graph.
The larger $\iter$, the larger and more complex motifs $\phi$ includes:
for small $\iter$, the kernel measures similarity with respect to simple
motifs, while for large $\iter$, similarity regards deeper, wider, and  more elaborate substructures.

A useful property of the WL kernel is that it can incorporate node attributes, which we refer to as \emph{node tags}. These can be cascade-specific (i.e., the same $v \in V$ can be tagged differently in $V_c$ and $V_{c'}$), and parameterized by {\em tagging functions} $\lblfunc_c: V_c \rightarrow \Sigma$, where $\Sigma$ is a finite set of attribute values. 
Importantly, to remain within our sanitized setting,
all tags will be derived either from the graph $G$ (e.g., node in-degree) or from the cascade $c$ (e.g., node depth), and not from any additional information external to the graph.
Furthermore, to prevent tags from inadvertently expressing node identities,
we use only coarse tags, where  node properties (e.g., in-degree)
are binned (typically logarithmically) into a small number of buckets.
Tags are useful in improving predictive performance.
For the remainder of the paper and when clear from context,
we will overload notation and assume $c$ also encodes node tags as determined by $\lblfunc_c$.

\subsubsection{Computing WL Representations}
We now describe the procedure for computing WL features
for a given cascade $c$ \citep{wlOriginal}.

The input consists of $c$, a tag-labeling function $\lblfunc$,
and a choice of the number of iterations $\iter$.

We proceed by iteration, with each indexed by $i$. Iteration $i$ associates a tag $\tg_i(v) \in \Sigma$ and a multiset of strings $M_i(v)$ for each vertex $v \in V$, where $\tg_0(v)$ is set initially to $\lblfunc(v)$.

In iteration $i$, we set $M_i(v) = \{\tg_{i-1}(v') | v' \in \mathfrak{N}(v)\}$, where $\mathfrak{N}(v)$ denotes the set of neighbors of $v$.

For each $v$, we sort and concatenate the strings $M_i(v)$ to obtain $s_i(v)$. Next, we prefix $s_i(v)$ with $\tg_{i-1}(v)$, the tag from the previous iteration, such that $s_i(v) := \tg_{i-1} || s_i(v)$, where $||$ is the concatenation operator. Last, we compress the new $s_i(v)$ by encoding it with a {\em hash function} $H: \Sigma^* \rightarrow \Sigma$, stipulating that $H(s_i(v)) = H(s_i(w)) \iff s_i(v) = s_i(w)$ (i.e., $H$ is a perfect hash function.\footnote{While this is theoretically impossible due to the pigeonhole principle, it is an easy condition for modern 32- or 64-bit computers to guarantee with near certainty. Indeed, our implementation simply hashes character string labels to an integer.}). We set $\tg_i(v) = H(s_i(v))$ for all $v$.

At each iteration $i$, the tag $\tg_i(v)$ of a node is thus a \emph{distinctive encoding} of a sequence of merges of tags from its neighbors. At iteration $i$, the tag for a node depends on information about vertices that are $i$ edges removed from the node. At the end of $\iter$ iterations, we compute $\#_i(c, \sigma_{ij})$, which is a count of the number of times the tag $\sigma_{ij} \in \Sigma$ occurs in the tags of $c$ at iteration $i$. Formally, let the set of tags associated with the vertices of $c$ at iteration $i$ be $\Sigma_i = \{\tg_i(v) | v \in V\}$. Assume without loss of generality that $\Sigma_i = \{\sigma_{i0}, \ldots ,\sigma_{i|\Sigma_i|}\}$ is lexically sorted. Then the mapping \\$\#_i: \Omega \times \Sigma \rightarrow \mathbb{N}$  represents the number of times that the tag $\sigma_{ij}$ occurs in $\Sigma_i$. Applied to $c$ and each tag in $\Sigma$, $\#_i$ induces a vector corresponding to the topological features of $c$:
\begin{align}
\phi(c)  = 
(\#_0(c,\sigma_{00}), \ldots, &\#_0(c,\sigma_{0 |\Sigma_1|}),
\ldots, \notag\\
& \#_\iter(c,\sigma_{\iter0}), \ldots, \#_\iter(c, \sigma_{\iter |\Sigma_\iter|}))
\end{align}%
That is, $\phi(c)$ is the concatenated values of the counts for each tag at each iteration, providing a topologically rich embedding of $c$.

\begin{figure}[t!]
	\centering
	\includegraphics[width=0.8\linewidth]{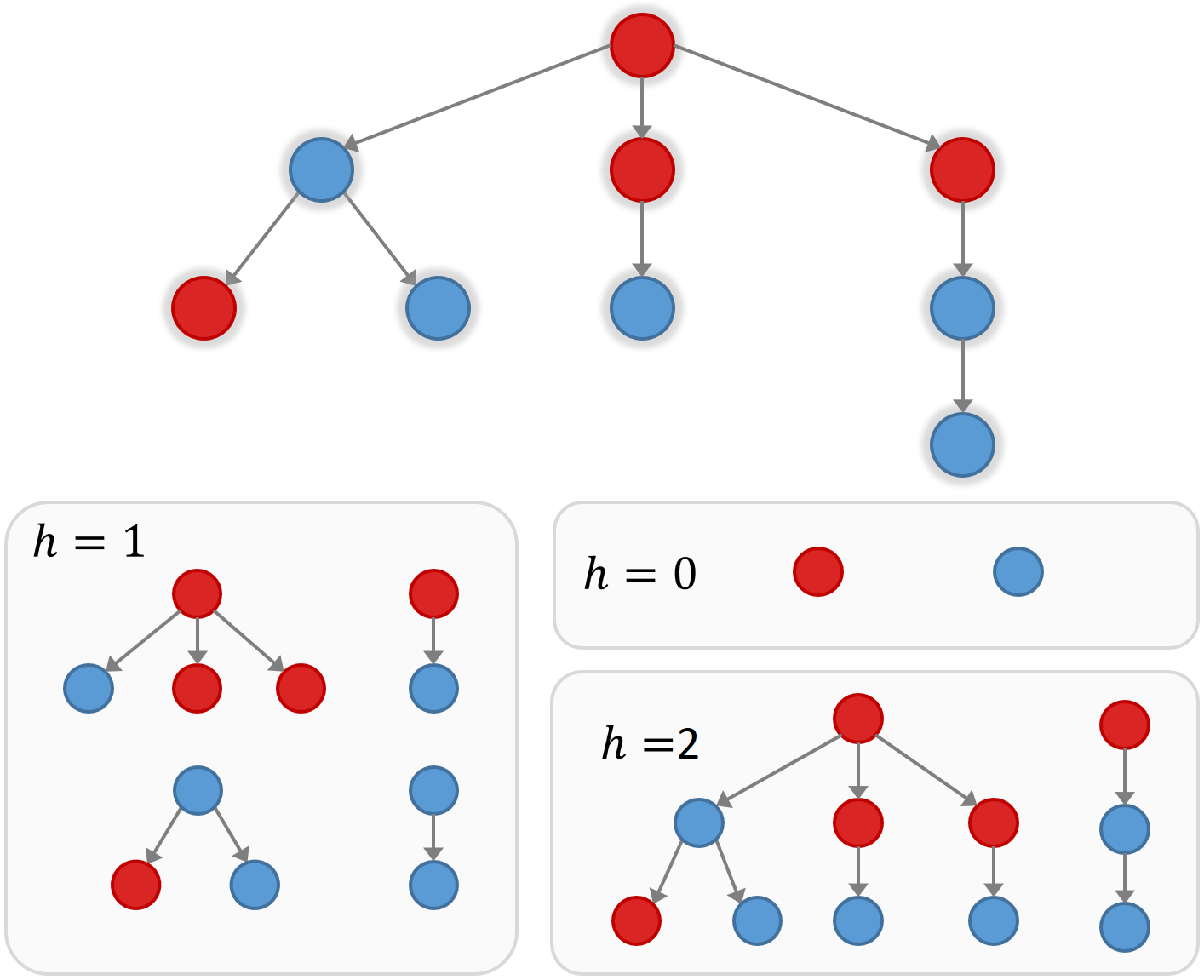}
	\caption{An illustration of the WL procedure. The cascade above is described as a collection of its sub-components,
		which each correspond to an entry in the feature vector $\phi$.
		Color represents tags $\tg \in \{\text{\texttt{\textbf{red, blue}}}\}$,
		which are taken into account in the construction of $\phi$.
		Boxes contain features obtained at iterations $\iter \in \{0,1,2\}$ of the WL
		procedure .}
	\label{fig:wl_illustration}
\end{figure}

\subsubsection{Runtime Analysis}
Let $m = |E_c|$ and $n = |V_c|$, noting that $m \geq n$. For one iteration of the WL kernel, each node receives and sorts its neighbors' tags. Because the list of all tags is finite and known, bucket sort can be used to order each node's multiset of neighbors' tags. There are $n$ buckets, one for each tag, and $m$ elements to sort, i.e., one tag passed along each edge. The sorting time is thus $O(n + m)$. The time to run $\iter$ iterations of the kernel is thus $O(\iter(n+m)) = O(\iter m)$.

\section{Experiments}
\label{experiments}

In this section, we present our experimental results in support of the hypothesis that patterns of information propagation can be informative of content. The experiments consider the problem of rumor veracity classification, where given a sanitized cascade of an unresolved rumor, the task is to predict if the rumor would eventually turn out to be true or false.

\subsection{Data}
\label{sec:data}
We evaluate our approach on use the dataset curated and studied by \citet{VosoughiRoyRumorsScience},\footnote{We note two considerations with respect to our use of this data.
	First, the data that we study are completely anonymous and only contain a few  non-identifying user-related attributes per node,
	and our work was done under approval of the Harvard University Institutional Review Board.
	Second, the dataset was provided to us with permission from Twitter, with whom we are not affiliated.
	Researchers interested in gaining access to the data should
	contact either \citet{VosoughiRoyRumorsScience} or Twitter directly.}
which includes Twitter retweet cascades linked to rumors that were verified and published by fact-checking websites.
Our main consideration in choosing a dataset was ensuring it could serve as a
credible testbed for evaluating our hypothesis.
The \citet{VosoughiRoyRumorsScience} dataset exhibits several unique characteristics, discussed below, that were crucial for our evaluation procedure.

\subsubsection{Data Collection and Processing}
\label{data_collection}
We begin by summarizing the process that \citet{VosoughiRoyRumorsScience}
undertook to collect veracity-labeled rumor cascades. 
Granted access to the full Twitter historical archive,
they began by searching for any tweet whose replies include a link to a fact-checking article from one of six websites.\footnote{These websites are snopes.com, politifact.com, factcheck.org, truthorfiction.com, hoax-slayer.com, and urbanlegends.about.com} They confirmed that the text of the fact-checking article was related to the content of the tweet by embedding both in a vector space and measuring their cosine similarity. They then collected all the retweets of the rumor, building the full cascade while discarding tweets that may have referenced the fact-checking article or were determined by a bot-detection algorithm \citep{varol_bot_detection} to have been written automatically. Last, they applied a method known as time-inferred diffusion \citep{goel2012structure} to reconstruct implied retweet edges by consulting the full Twitter follower graph, an index of the relationships among all Twitter users that is not made available to the public.
The label of each cascade, i.e., a binary value $y$ reflecting whether the rumor turned out to be true or false,
was set in accordance with the veracity of the cascade as revealed by the relevant fact-checking organization.

Critically, \citet{VosoughiRoyRumorsScience} undertake rigorous postprocessing steps
to ensure that all events in each cascade precede the publication of the corresponding fact-checking article. The curation process results in a collection of veracity-labeled rumor cascades whose participating users 
are likely \emph{unaware of any publicly available resolution regarding the veracity of the rumor at their time of sharing}.
This supposition is further strengthened by the inclination of fact-checking websites to target and publish results on nontrivial rumors whose resolution requires time and effort.
Additionally, since the dataset was constructed from the full Twitter archive,
cascades are fully representative of Twitter user dynamics
and do not suffer from issues of partiality or bias due to sampling restrictions characteristic of most other Twitter datasets.
To the best of our knowledge, the \citet{VosoughiRoyRumorsScience} dataset is unique in having these properties.
For further details, we refer the reader to the supplementary material in their work.

Note that due to the retweet mechanics of Twitter and to the nature of our dataset, cascades represent only partial observations. 
Specifically, while we can observe a post (a tweet or retweet) by a user,
we cannot know which of their followers saw but did not retweet
that post. This means that leaf nodes in the cascades represent users who posted some content but whose followers did not retweet that same content within the observed time period.

\subsubsection{Data Characteristics}
\label{exp:data_characteristics}

In total, the full dataset contains about 126,000 cascades comprising over 4.5 million tweets. Both cascade size and depth distributions are heavy-tailed, 
as is typical in such data.
To avoid trivial cascades, in our experiments we only consider cascades of size 25 or more nodes, which account for 4\% of all instances.
Overall, we experiment on 5,066 cascades.
Label classes are imbalanced, with true rumors accounting for only 21\% of all rumors and 14\% of all cascades. Our choice of evaluation metrics and optimization approaches, described later, account for this.

The cascades used in our experiments are sanitized in three important ways.
First, they do not include any textual information (this is not available in the  dataset).
Second, they do not convey any notion of time (retweet time signatures are available in the data, but we discard them).
Third, they do not reveal any user identities.
While the original data are anonymous, users could still in principle be linked across cascades through other identifiers (e.g., the unique degree of a popular node).
However, our use of log-scaled bins for node tags ensures that any inferred attribute is shared by many nodes, making individual identification extremely unlikely.
Moreover, the variance in retweeting patterns across cascades
provides a natural ``masking'' of local neighborhoods, making
linkage across cascades via graph isomorphism techniques essentially infeasible.


\subsubsection{Limitations}
\label{exp:data_limitations}

As with any data collection procedure, the dataset used by \citet{VosoughiRoyRumorsScience} presents certain limitations that are important when considering conclusions and implications of our results.

First, the dataset includes only rumors that were deemed worthy of investigation and publication by fact checking websites, and only cascades that directly referenced those published results were associated with a rumor. Hence, the data does not consider all cascades emanating from a specific rumor, the entire population of potentially-rumorous cascades, or the general set of all Twitter cascades. 
Misinformation is notoriously difficult to define, making its collection and labeling challenging and susceptible to misinterpretation or  criticism.
The approach taken by \citet{VosoughiRoyRumorsScience}, and which we adopt herein,
is to appeal to  third-parties, such as fact-checking organizations that
effectively define and label a particular type of misinformation, namely online rumors.
While this may introduce a measure of selection bias, it is hard to imagine a way to collect potentially misinformative content that does not introduce some bias.
Nonetheless, to further validate their results,
\citet{VosoughiRoyRumorsScience} repeated their experiments with rumors identified and labeled by independent labelers (that were not considered by fact-checking websites), finding that their results held.

We further note that content that has been targeted, researched, and published by fact-checking websites likely consists of interesting content useful for evaluating our method.

Second, despite the extensive efforts of \citet{VosoughiRoyRumorsScience} to control for possible confounding factors, information regarding content can still leak into cascades, (e.g., from external sources of information), which is impossible to mitigate fully. However, the successful replication of their findings on the set of rumors identified and labeled independently provides evidence against such confounding influence.

Third, the dataset cannot (and should not) be considered a collection of fake news. Fake news is a term that disregards important factors such as malintention, publication medium, and the potential for news to have mixed veracity. It does not distinguish between maliciously false news and satire, it cannot be objectively identified, and its evolution in the public vernacular has resulted in its use by some politicians as a catchall for opposing positions or unflattering portrayals. 
This noted, clearly there is a connection between the spread of rumors and the phenomenon of fake news, and our results should be relevant to research in this domain.

While the dataset serves our purpose of determining veracity based on patterns, it is not ideal for evaluating the usefulness of predictive methods as real-time classification methods, while rumors spread. This is because these data are collected after the fact,
when the veracity of the rumor has already been ascertained and publicly announced. What we are able to do is to simulate a real-time setting by truncating cascades by time and investigate how early in a rumor's diffusion our method can accurately classify veracity. Our results in this setting, described in Section \ref{exp:truncation_time}, are encouraging.

Finally, we note that the above issues are intrinsic properties of the domain of rumor propagation in social networks. To the best of our knowledge, the dataset that we use is the best currently available source of data on this topic.

\subsection{Experimental Setup}
\label{exp:setup}

We focus on cascades composed of at least 600 nodes,
a number chosen to strike a balance between having sufficiently rich cascades and enough examples to learn a robust classifier (see later discussion in Section \ref{sec:min_size}).
For our method based on WL kernels, we consider four variants, 
differing in predictive models and types of tags used.
The predictive models include a \emph{linear} model and a \emph{nonlinear} model,
while tags are based on either the \emph{cascade} or the underlying Twitter follower \emph{graph}.
We  refer to these methods as \emph{WL-lin-c, WL-lin-g, WL-nonlin-c},
and \emph{WL-nonlin-g},
and use the terms \emph{WL methods} (or simply \emph{WL} when clear from context) to refer to any or all of them.
For the linear model, we use logistic regression with L2 regularization. For the nonlinear model, we use gradient-boosted trees (using the LightGBM package \citep{lgbm}), which are known to be especially performant when applied to sparse representations such as those of the WL kernel.

For node tags, we use the out-degree value:
for cascade-based models, these are computed with respect to the cascade,
and the graph-based models, they are computed with respect to the underlying Twitter follower graph.\footnote{The dataset provides only in- and out-degree counts for each node. It does not provide direct access to the Twitter follower graph.}
Note that the outgoing edges in the former are an instance-dependent subset of the outgoing edges in the latter.
As discussed in Section \ref{model_wl}, rather than assigning actual degree values to nodes, we assign node tags based on the logarithmically-scaled bin a node's out-degree falls into. We bin values both to reduce the effective dimensionality of the feature representation and to mitigate the possibility of inadvertently identifying users on the basis of unique degree values.

To prevent label information from leaking across cascades,
we execute a train-test split that is stratified by rumor. That is, if a rumor is associated with several cascades, those cascades will all be in either the train or test set.
To account for the class imbalance in our dataset,
in all experiments we report the F1 score as a metric of predictive performance.
We repeat each experiment 100 times with different random seeds and report mean results across trials.

Tuning across regularization constants, penalty functions, and any other hyperparameters for all models was done using 5-fold rumor-stratified cross validation on the training set.
The parameter space was explored by Bayesian optimization.

\begin{figure*}[t!]
	\includegraphics[width=0.8\linewidth]{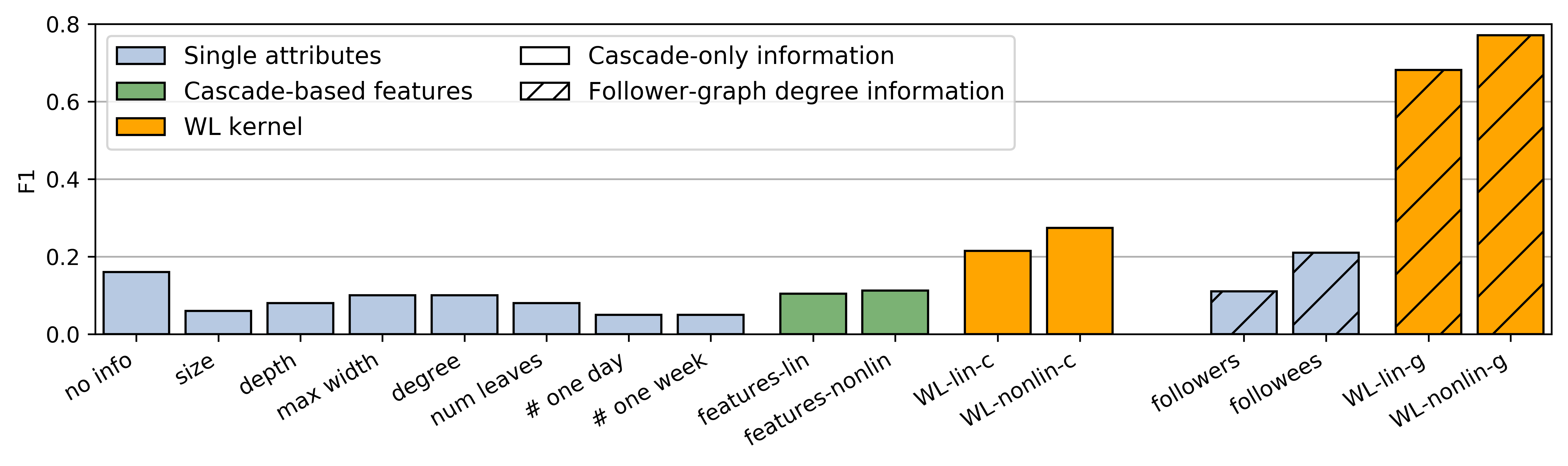}
	\caption{Predicting the veracity of online rumors based on sanitized Twitter cascades.
		Single attribute baselines (\textit{attributes})
		and graph feature-based linear (\textit{features-lin}) and nonlinear (\textit{features-nonlin}) classifiers 
		 perform on par with
		(or worse than) a no-information (\textit{no info}) baseline (i.e., a random Bernoulli predictor parameterized with the optimal class balance).
		WL methods significantly outperform all baselines
		and perform exceptionally well when nodes are tagged with their
		log-binned follower graph out-degree counts,
		with the nonlinear WL model (\textit{WL-nonlin-g}) achieving $\text{F1} = $ 0.77, improving over the no-information baseline by a factor of 4.81.}
	\label{fig:main}
\end{figure*}

\subsubsection{Baselines}
\label{exp:baselines}
One of the central contentions of this work is that standard graph metrics, even if known to differ significantly between true and false rumor cascades, do not hold sufficient predictive power to accurately discern veracity. In contrast, the WL kernel, which represents a cascade as a large collection of its substructures, offers a data-driven approach to determining which aspects of cascade topology are important for predictive accuracy. To test this claim, we employ baseline models that use cascade attributes shown to correlate strongly with veracity \citep{VosoughiRoyRumorsScience} and graph metrics used successfully in other cascade classification tasks \citep{ratkiewicz2011detecting}. We compare the predictive power of these baselines to that of the WL methods such that the difference in performance, if any, provides quantitative evidence as to the utility of the WL representation.

Overall, for each cascade we compute 32 handcrafted baseline graph attributes.
Simple cascade attributes include size, density, depth, width,
number of leaf nodes, and out-degree of the root node.
We include several summary statistics for degree and centrality distributions, as well as assortativity values.
We complement these with attributes describing non-topological aspects of the cascade: the number of nodes retweeting within one day and within one week of the root tweet,
and the structural virality of the cascade \citep{VosoughiRoyRumorsScience}.

We use these attributes in two ways:
individually and as input to supervised classifiers.

The purpose of the individual attribute baselines is to demonstrate the low predictive power inherent to these attributes, despite their established correlation with veracity \citep{VosoughiRoyRumorsScience}.
For a given attribute, predictions are generated by first learning from a training set a threshold that maximizes F1 score, then predicting as false any cascade whose attribute value is below the threshold.
We report attribute baselines for {\em  cascade size, depth, maximal width, root degree, number of leaves, number of tweets after one day, number of tweets after one week, median followers, and median followees}.\footnote{Though our WL methods use tags based on binned out-degree, our attribute baselines consider the unbinned median number of followers and followees of users involved in the cascade. Note that here an unbinned attribute baseline is a weakly better model than a binned attribute baseline.}

We then combine these individual attributes into vector inputs for trained classifier models. These stronger baselines aim to quantify the added predictive power afforded to a model by the WL kernel. Therefore, we use the same kinds of classifiers as in the WL methods, namely logistic regression (\textit{features-lin}) and gradient-boosted trees (\textit{features-nonlin}).

For completeness, we also compute a no-information baseline (\textit{no info}), which we define as the highest-achievable F1 score when no input information is available (akin to comparing to chance when evaluating with respect to accuracy).
If cascades are uninformative of content, all predictive models given cascades as input would perform as well as or worse than the \textit{no info} baseline. However, if cascades are informative, then a model's outperformance of \textit{no info} provides insight into the degree of their informativeness.

We select baselines conducive to an appropriate experimental environment for testing our hypothesis. We choose not to include baselines that draw on more recent approaches such as deep graph networks (e.g., \citet{yanardag2015deep})
for three reasons.
First, such deep methods tend to require vast amounts of labeled data, while the number of cascades in our dataset is small by those standards.
Second, there is increasing evidence that such methods lack robustness \cite{zugner2019certifiable},
which does not align well with part of the motivation for this work.
Third, and most importantly, this work seeks to test a hypothesis, not achieve state-of-the-art predictive performance. 
To this end, it suffices to demonstrate that some suitable predictive method over sanitized cascades is sufficiently accurate.
As we show, we are able to validate our hypothesis with our chosen kernalized approach.

Finally, we note that an ideal set of baselines would include methods that consider content (e.g., retweet text) or user identities. Unfortunately, the dataset does includes neither text nor any kind of user identifiers, making such comparisons impossible.

\subsection{Experimental Results}

We now present our main results for predicting rumor veracity from sanitized cascades (Figure \ref{fig:main}).
We begin with methods that use information available only in sanitized cascades.
The results show that simple attributes of a cascade (\textit{attributes}) are not powerful predictors. In fact, the F1 scores associated with these attribute-based baselines are similar to the score of the no-information baseline (\textit{no info}) with $\text{F1}=0.16$,
suggesting that these models are incapable of extracting from cascades information that is relevant to their veracity. Graph feature-based baseline classifiers (\textit{features-lin} and \textit{features-nonlin})
do not fare much better, regardless of the underlying predictive model. Meanwhile, graph kernels methods with cascade-based tags (\textit{WL-lin-c} and \textit{WL-nonlin-c}) 
achieve F1 scores significantly higher than the no-information baseline,
suggesting that they are able to discover veracity signals from sanitized cascades alone.

While demonstrating informativeness, the predictive power of cascade-only WL methods
remains fairly low, with $\text{F1}=0.25$ for \textit{WL-lin-c} and $\text{F1}=0.27$
for \textit{WL-nonlin-c}.

However, the WL scores improve dramatically when node tags correspond to log-binned follower graph out-degrees, known on Twitter as the number of \emph{followees}.
The single attribute baselines suggest that when considered alone, information drawn from the follower graph (such as the number of followers and followees) provides limited predictive power.
But WL methods are able to leverage such signals in the context of the cascade's broader topology to produce surprisingly accurate predictions,
achieving  $\text{F1}=0.68$ for \textit{WL-lin-g} and $\text{F1}=0.77$
for \textit{WL-nonlin-g}.
Hence, injecting into the rich WL structural cascade representation a small amount of anonymous, content-blind, and very coarse local graph-based information results in accuracy that outperforms the no-information baseline by factors of $4.25$ and $4.81$ for the linear and nonlinear models, respectively.

This result embodies our most important finding,
that cascade topology is indeed informative of rumor veracity.
Next, we analyze the predictive properties of graph kernels, investigating their performance under two different experimental conditions and revealing interesting tradeoffs.

\subsubsection{Varying Minimum Cascade Size}
\label{sec:min_size}

In the first experimental condition, we evaluate the performance of graph kernel methods
on cascades of varying sizes (by number of nodes).
Specifically, for increasing values of size threshold $m$,
we test how well graph kernel methods perform when restricted to cascades whose size is at least $m$. 

Our goal in this experiment is motivated by practical concerns:
intriguing rumors are likely to be discussed more
and therefore lead to larger cascades. Hence, such large, socially impactful cascades are of prime interest in terms of accurate prediction.
Moreover, as they contain more information,
it is plausible that they are easier targets for prediction than small cascades.
Therefore, we may expect that setting a high threshold value (thus including in the sample set only large cascades) will improve performance.

On the other hand, large cascades are also considerably rarer in our dataset, so increasing the threshold will result in smaller training sets. As such, we might also expect performance to degrade with an increasing threshold.
Here we explore this tradeoff and ask whether filtering out smaller cascades improves overall performance, and if so, to what degree.

\begin{figure}[t!]
	\includegraphics[width=\linewidth]{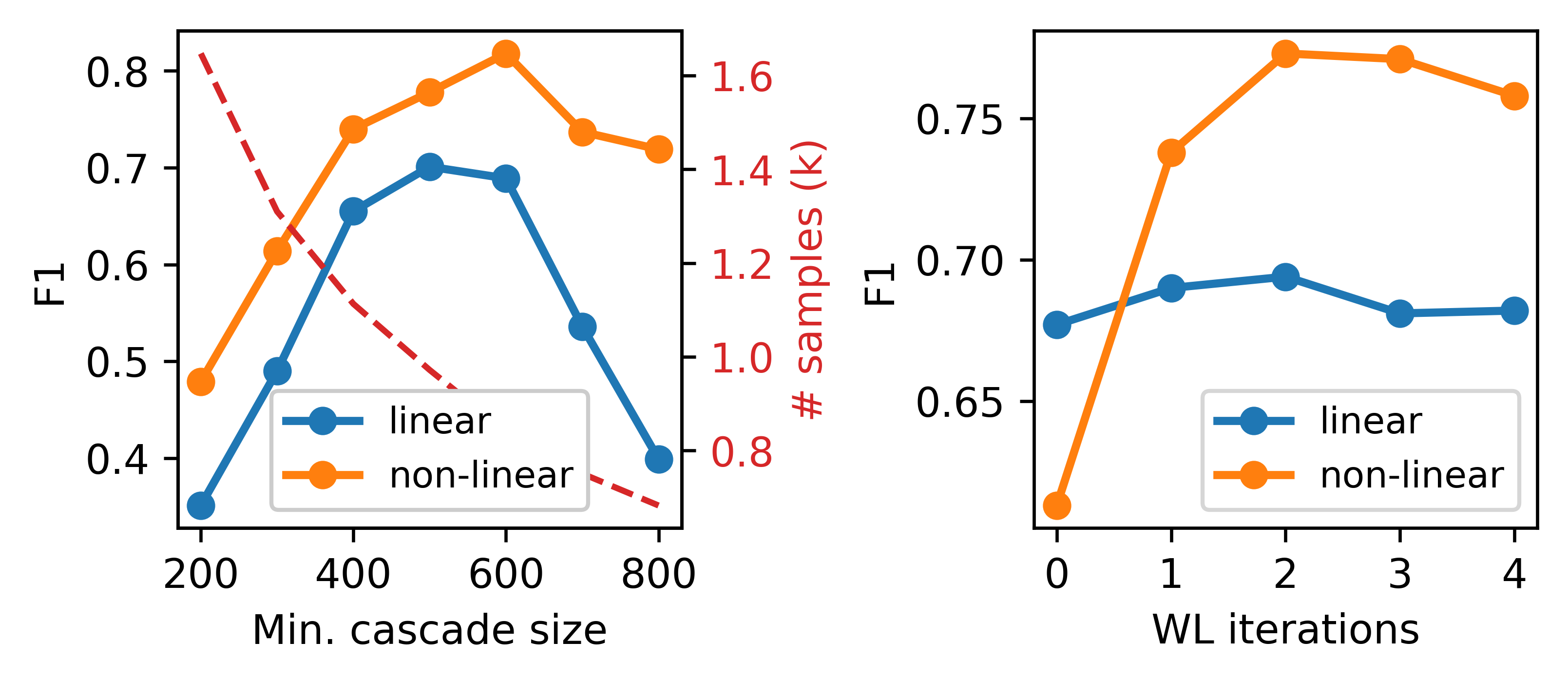}%
	\caption{Performance of WL when varying cascade size and number of WL iterations.
		\textit{Left}: Performance of WL on data with increasing minimal cascade size (i.e., number of nodes). When the threshold increases, the data include larger and hence more informative cascades, but the number of data points decreases. This results in an inverse U-shaped curve. The nonlinear WL is more robust to a decrease in the number of large cascades.
		\textit{Right}: Performance of WL for an increasing number of WL iterations ($\iter$). As the number of iterations increases, cascades representations consider richer substructures.
		The nonlinear model utilizes these expressive representations to improve predictive performance.
		Meanwhile, the linear model is fairly insensitive to the number of iterations,
		indicating that it is unable to leverage higher-order graph substructures,
		and that modeling nonlinear interactions between topological features is necessary for achieving good accuracy.
	}%
	\label{fig:size_and_h}%
\end{figure}

Figure \ref{fig:size_and_h} (left) shows F1 scores for threshold $m \in \{200,300,\dots,800\}$,
as well as the number of samples for each threshold.
As can be seen, performance across thresholds reveals a pattern of poor performance at extreme values of $m$; setting the threshold too low leads to many non-informative cascades,
while setting it too high leads to insufficiently few cascades in the training set.
Based on this, we select 600 as our canonical cutoff
for all other experiments in which we fix the minimal cascade size.

\subsubsection{Varying WL Iterations}
In our second experimental condition, 
we test the performance of the WL graph kernel method as the number of WL iterations increases. Recall that at iteration $\iter$, graph kernels represent each cascade by substructures
that are of depth at most $\iter$.
Increasing the number of iterations has two contrasting effects.
On one hand, increasing iterations leads to a representation of cascades that is more nuanced and informative. Hence, performance could potentially improve.
On the other, having deeper substructures also means having more
substructures. As a result, the effective number of features per cascade increases,
and overfitting may cause performance to degrade.

Figure \ref{fig:size_and_h} shows results for $\iter \in \{0, 1,2, 3, 4\}$.
The results reveal a stark distinction between the linear and nonlinear models.
For the nonlinear model, more iterations are in general helpful
and contribute significantly to performance:
the optimal score at $\iter=2$ is 0.77, whereas for $\iter=0$
(where substructures include only node-type counts) it is 0.68.
Mild overfitting can be observed starting at $\iter=3$, suggesting that performance could further increase with access to larger datasets.
For the linear model,
the performance curve is flat,
indicating that the classifier is incapable of utilizing the 
additional information available in higher-order structures.
Taken together, these results imply that educing predictive power
from higher-order substructures requires modeling their interactions.

\begin{figure}[t!]
	\includegraphics[width=\linewidth]{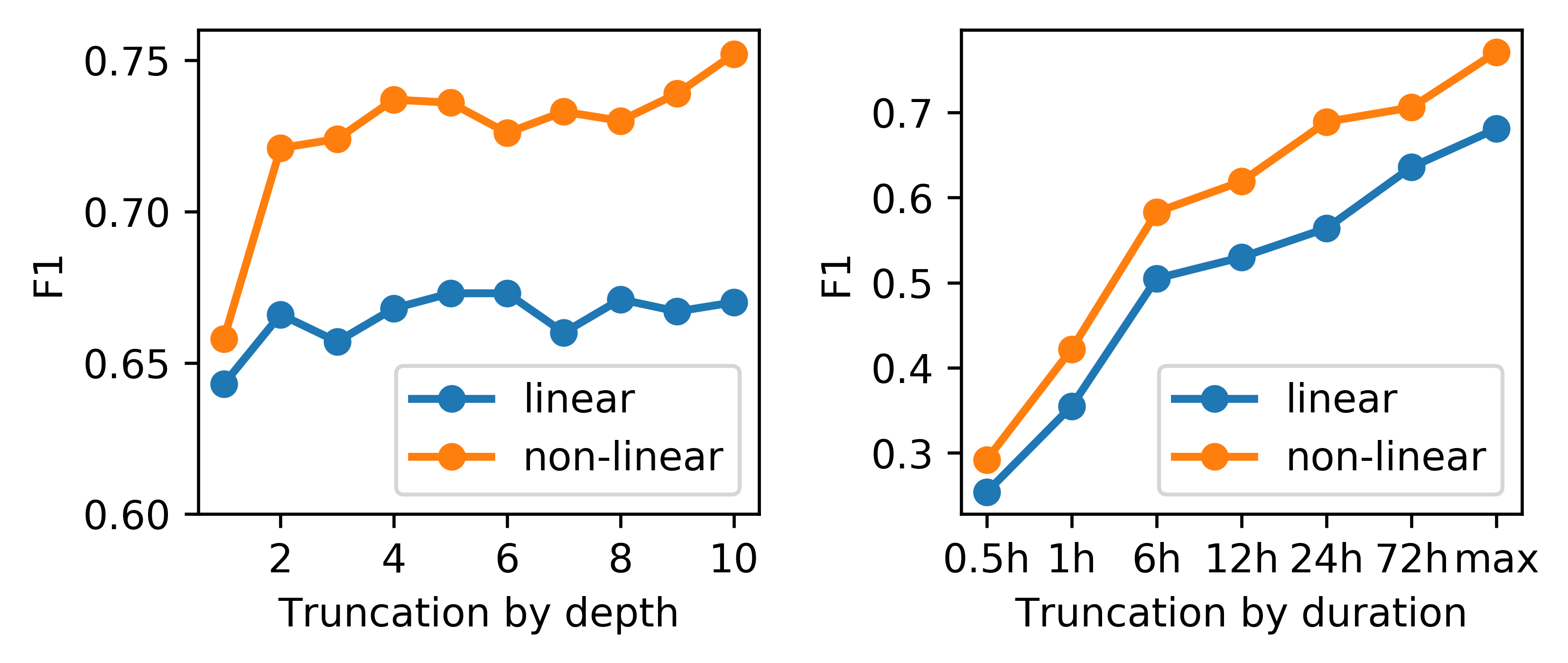}
	\caption{Performance of WL on cascades truncated by depth
		and duration.
		\textit{Left}:
		For depth, most of the increase in performance occurs when considering all nodes at distance at most two from the root (i.e., children and grandchildren). Nonetheless, adding deeper nodes continues to improve performance, an effect most salient in the nonlinear model.
		\textit{Right}:
		For duration, performance steadily increases as more information accumulates over time.
	Further, both linear and nonlinear WL models, when given data gathered up to only one hour from the initial posting time, outperform all baselines given access to all of the data. By 24 hours, performance of the nonlinear WL model reaches 
		roughly 90\% of the score it achieves when trained on the full data.
	}
	\label{fig:depth_and_time}
\end{figure}

\subsection{Analysis}
\label{exp:analysis}

So far, we have considered the cascades in the data as static artifacts. In reality, cascades are dynamic objects that evolve over time. One implication of the data collection process (see Section \ref{sec:data}) is that each instance represents a cascade as it was observed just before its rumor's veracity was publicly disseminated. Hence, the main experiment's setup assumes the viewpoint adopted at the peak of each cascade's informativeness relative to what is in the public domain.
However, investigating whether cascades are indicative of veracity also demands studying predictiveness when the observations used to classify rumors are made earlier in the development of a cascade.
To this end, we consider two measures of the progression of a cascade. The first is \emph{time locality}, or the duration from the original tweet, and the second is \emph{graph locality}, or the number of edges from the root node in the cascade.

\subsubsection{Truncation by Time}
\label{exp:truncation_time}

The first variation simulates an early detection scenario, where cascades are observed only a short time after the initial tweet. For this, we instrument our dataset such that each cascade is replaced by a \emph{time-truncated subgraph} of the cascade. A cascade truncated at a duration of $t$ hours contains only nodes of that cascade that correspond to retweets posted within $t$ hours of the root tweet. In this way, the  truncation gives the cascade that would have been observed had the measurement been taken $t$ hours after the root tweet.

Our goal here is twofold. Beyond simply understanding the reliability of the method at early stages in the propagation process, demonstrating successful early prediction also alleviates possible concerns of leakage with respect to any publicly disseminated signals about the veracity of a rumor. Leaf nodes of the cascade appear on average 11 days after the root tweet. Hence, the earlier in a cascade we can show that that the structure of diffusion is predictively useful, the more unlikely it is that our models are learning from topological signals that represent latent fact checking information.

Figure \ref{fig:depth_and_time} presents results for a variety of different truncation levels.
Our findings show that, even within one hour of the original tweet, the WL methods surpass all other baselines considered in Section \ref{exp:baselines}, even though these baselines are given the full, untruncated cascades.
Within 24 hours of the root tweet, the nonlinear WL model achieves 89.4\% of the performance without time-based truncation, and within 72 hours it matches 91.6\%. Truncating at one hour and 24 hours results in cascade sets containing just 17\% and  57\% of all nodes, respectively. These results demonstrate that veracity signals can be captured early in the propagation process.

\subsubsection{Truncation By Depth}

As a second kind of robustness check, we truncate cascades by the distance of nodes from the root tweet. That is, for some depth $d$, any node more than $d$ edges from the root is removed from the cascade. As before, this process results in valid cascades that are pruned radially from the root.
Our interest here is to explore a setting in which the nodes in the neighborhood of the root tweet may be considered trustworthy and part of an established community. Perhaps these trustworthy root nodes are {\naive}ly sharing questionable content, while deeper nodes in the cascade behave adversarially in accordance with an ulterior motive, such as the spread of misinformation. Accurate prediction on cascades that are restricted to small neighborhoods around the root tweet would imply that malicious actors would need to find ways to exercise influence from deep within established communities--- a much harder task.

Figure \ref{fig:depth_and_time} presents results at different truncation depths. While the largest increase in performance can be attributed to the inclusion of the root's grandchildren (i.e., depth one vs. depth two nodes), the model's performance continues to improve as deeper substructures are considered. Interestingly, these results also shed light on which of the underlying graph substructures are important for high accuracy. As cascades become deeper, the WL process collects additional subgraph signatures, especially so when the number of WL iterations is larger. Hence, the results suggest that features extracted from complex motifs are crucial for properly aggregating the crowd's signal. Note also that this is only apparent in the nonlinear model. In the  case of the linear model, adding depth does not dramatically affect performance, a fact that aligns well with our findings with respect to WL iterations (Figure \ref{fig:size_and_h}). Hence, not only are individual deep structures important, but so too are their interactions, which are vital to the accurate prediction of veracity.

\subsubsection{Interpretability}
One of our motivations for including a linear WL classifier was that, when trained with a sparsity-inducing regularizer like LASSO, the model provides a framework for interpretation that could potentially shed light on the predictive importance of individual features (i.e., cascade substructures or motifs).
This, however, proved to be challenging.
As our results show, the linear model is insensitive to the number of WL iterations
(see Section \ref{sec:min_size}).
Further experiments with sparse linear models revealed two related phenomena.
First, training results in features selected inconsistently (i.e., different random train-test splits produce models with similar accuracy that rely on significantly divergent sets of features). Second, in many cases the selected features correspond to $\iter=0$.

Furthermore, the superior performance of the nonlinear model suggests that much of its predictive power stems from considering interactions between features,
making interpretation all the more challenging.
Hence, our analysis shows not only that linear models provide little insight as to the predictive mechanics of graph kernels on our task, but also that merely considering individual motifs, rather than their interactions, is insufficient for good accuracy.

\section{Discussion}

In this work, we consider whether network diffusion patterns are informative of content. By building models that learn representations of sanitized cascades, we show that strong predictive performance of rumor veracity can be achieved without considering any textual or user-identifying information.
In particular, we demonstrate that topologically rich encodings of cascades provide enough information to predict veracity.

These experimental results support the hypothesis that aspects of content, likely opaque even to individual participants, can be accurately predicted from the collective signal of diffusion patterns.
At the same time, they also suggest that accurate prediction requires careful consideration as to how cascades are represented. 

In our work, we represent rumor cascades as collections of small, high-fidelity graph substructures. The models that we train are able to efficiently aggregate these motifs of communication flow for prediction, and the method as a whole can be considered a flexible, data-driven approach for eliciting the wisdom of the crowd with respect to a rumor's veracity.
We choose to use structures derived from the WL kernel, 
a powerful and general approach for embedding graphs.
We hope that future research will reveal additional cascade representations that are both useful and interpretable.

We also observe that unlike content-based approaches that analyze the authors, text, images, and metadata associated with individual tweets, the proposed model aggregates simple behavioral signals of many individual users. We argue that this provides the additional benefit of {\em community robustness}: for a malicious producer of misinformation to succeed in fooling our topological models, they must influence the actions of a large number of honest users.
The extent to which this sort of robustness can be demonstrated empirically is limited by the data that are currently available to the research community. While recent work has focused either on the brittleness of content-based models \citep{belinkov2019analysis} or  the propensity of graph-based predictors to thwart adversarial attacks \cite{zugner2019certifiable},
it does not consider them jointly. Such research would be illuminating, but to the best of our knowledge, there are no available datasets in which both modalities are present. We hope that the results herein encourage the curation and release of such data.

We conclude by noting that although our research may enrich the body of work on the prediction of misinformation, real impact on the future of the Web requires intervention. One way for policymakers to limit the spread of misinformation is to dictate global rules and enforce them by removing false content and banning users. But such harsh measures are controversial. Who adjudicates what constitutes false content or who is a toxic user? The approach outlined in the present paper sidesteps these questions: we show that the local actions of individual users convey information regarding the veracity of the rumors with which they engage. Hence, platform-wide outcomes might be shifted not by imposing a set of top-down rules, but by incentivizing users to act individually in ways that are shown to be globally helpful. Of course, determining how to guide users at the local level requires a strong causal understanding of online information propagation, which is the ultimate goal when studying the role of the Web in shaping our social discourse.



\begin{acks}
We would like to thank our anonymous reviewers for their constructive comments. We are grateful to Deb Roy and Soroush Vosoughi for the use of their dataset.
\end{acks}

\bibliographystyle{ACM-Reference-Format}
\bibliography{refs}

\appendix


\end{document}